\documentclass[12pt, a4paper]{article}
\usepackage[font=footnotesize]{caption}
\usepackage[utf8]{inputenc}
\usepackage{graphicx}
\usepackage{authblk}
\usepackage{booktabs}
\usepackage{verbatim}
\usepackage{url}
\usepackage{caption}
\usepackage{subcaption}
\usepackage{indentfirst}
\usepackage{url}
\usepackage{amssymb}
\usepackage{amsmath}
\usepackage{tcolorbox}
\title{ Agglomerative Clustering in \\ Uniform and Proportional Feature Spaces}

\author{Alexandre Benatti$^1$ and  Luciano da F. Costa$^2$}

\affil{
$^1$Institute of Mathematics and Statistics - DCC \\
University of S\~ao Paulo \\
Rua do Mat\~ao, 1010, \\ S\~ao Paulo, SP 05508-090 Brazil 
\\ \vspace{0.5cm}
$^2$S\~ao Carlos Institute of Physics - DFCM \\
University of S\~ao Paulo \\
Av.~Trabalhador S\~ao-carlense, 400, \\ S\~ao Carlos, SP 13566-590 Brazil
}

\date{\emph{10th April, 2024}}

\begin{document}

\maketitle

\begin{abstract}
Pattern comparison represents a fundamental and crucial aspect of scientific modeling, artificial intelligence, and pattern recognition. Three main approaches have typically been applied for pattern comparison: (i) distances; (ii) statistical joint variation; (iii) projections; and (iv) similarity indices, each with their specific characteristics. In addition to arguing for intrinsic interesting properties of multiset-based similarity approaches, the present work describes a respectively based hierarchical agglomerative clustering approach which inherits the several interesting characteristics of the coincidence similarity index --- including strict comparisons allowing distinguishing between closely similar patterns, inherent normalization, as well as substantial robustness to the presence of noise and outliers in datasets. Two other hierarchical clustering approaches are considered, namely a multiset-based method as well as the traditional Ward's approach. After characterizing uniform and proportional features spaces and presenting the main basic concepts and methods, a comparison of relative performance between the three considered hierarchical methods is reported and discussed, with several interesting and important results. In particular, though intrinsically suitable for implementing proportional comparisons, the coincidence similarity methodology also works effectively in several types of data in uniform feature spaces. 
\end{abstract}

\section{Introduction}\label{sec:introduction}

Conceiving, validating, and applying \emph{models} for better understanding and predicting natural phenomena constitute the main activities underlying \emph{Science} and the scientific method (e.g.~\cite{da2019modeling}). At the same time, humans, as well as several other living beings, rely on \emph{pattern recognition} for survival and perpetuation. Pattern recognition (e.g.~\cite{fukunaga2013,bishop2006,duda2000pattern,da2018shape}), which can take place in a supervised (e.g.~\cite{amancio2014systematic}) or non-supervised (e.g.~\cite{rodriguez2019clustering}) manner, consists of identifying and assigning categories to patterns. Indeed, despite our limited memory and computational abilities, humans have excelled in several pattern recognition tasks to the point that only more recently automated approaches became capable of approaching our performance respectively to identifying patterns of several types.

Interestingly, modeling and pattern recognition can be ultimately conceptualized as being closely interrelated one another, or even constituting the same basic activity (e.g.~\cite{da2019modeling}). Briefly speaking, the construction of models relies on the preliminary identification of repetitive (e.g.~an oscillating pendulum) or atypical (e.g.~a possible new species) patterns, which motivate respective systematic studies involving their categorization and modeling. Remarkably, despite the special importance of pattern recognition for humans, the implementation of this full ability into machines has represented a difficult challenge for science and technology.

Henceforth, we understand as \emph{pattern} any signal of special relevance or interest. In addition, the selection of the variables and parameters to be considered in a model can also be understood from the perspective of feature selection. The results obtained from model simulations also require them to be contrasted with real-world respective signals, demanding the important ability to \emph{compare} patterns. At the same time, recognizing a pattern requires it to be preliminary \emph{compared} with other signals, which allows it to be modeled in terms of examples, prototypes, or respective sets of rules, which can then be \emph{compared} with new data.

It follows from the above considerations that the ability to \emph{compare} two patterns stands out as being both central and critical to \emph{both} scientific modeling and pattern recognition. Basically, comparing two patterns can be understood as quantifying how much they resemble (or differ) one another.   Though there is a virtually unlimited number of possible manners in which two patterns can be compared, humans and machines have mostly considered in terms of the three following possibilities: (i) \emph{distances} between patterns; (ii) statistical \emph{joint variation} between patterns; (iii) projections; and (iv) \emph{similarity} between patterns.  

Given a pattern of any type, its objective comparison and identification require it to be first mapped into a finite set of respective \emph{measurements}, \emph{properties}, \emph{characteristics}, or \emph{features}, which are often organized as a respective \emph{feature vector} which can be represented in a respective \emph{feature space}. For instance, an apple can be represented by a respective three-dimensional feature vector $\left[weight, size, sugar content \right]$. It can be soon realized that several such vectors are possible, depending on the choice of measurements to be taken into account. Actually, the choice of suitable feature vectors constitutes the first challenge involved by pattern recognition. The choice of a particular set of features defines the \emph{feature space} where the respective feature vectors can be represented. 

Pairwise relationship between two data elements or subclusters constitutes the basis of several approaches to pattern recognition, including a family of methods known as \emph{agglomerative hierarchical clustering} (e.g.~\cite{ward1963hierarchical,duda2000pattern,da2018shape,murtagh2012}). These methods are particularly interesting because they do not require previous knowledge about the number of clusters and provide a complete indication of the relationships between clusters and subclusters along successive hierarchical levels, which can be effectively represented and visualized in terms of respective dendrograms. The present work describes two hierarchical clustering approaches based on multiset similarity indices, especially the \emph{coincidence similarity index} (e.g.~\cite{da2021further,costa2022simil,da2022coincidence,costa2023mneurons}). A related approach also employed the coincidence similarity for hierarchical clustering was previously described in~\cite{da2021real}, but in that case, the comparisons took place between the densities of existing subclusters.

Another important topic receiving special attention in the current work concerns the concept of uniform and proportional feature spaces, which are described, illustrated, and employed for the comparison between three agglomerative hierarchical clustering methods, namely two approaches based on multiset similarity as well as the traditional Ward's methodology. These methods have their performance compared in terms of a respective index while considering three distinct types of clustering structures in two-dimensional spaces.

The present work starts by briefly discussing the three main approaches that can be adopted for comparing two data elements, and then proceeds by describing the concepts of uniform and proportional features spaces, as well presenting the adopted similarity indices and describing the considered models of cluster in both uniform and proportional feature spaces. Experimental results involving several types of clustering structures in both types of feature spaces are then reported and discussed.

\section{Four Comparison Approaches}

A great deal of patterns treated by humans have a visual/geometric nature, being henceforth referred to as \emph{shapes} that are intrinsically assumed to be invariant to translation, rotation, reflection, and often magnification. For instance, a pen remains a pen irrespectively of these three transformations. These types of patterns have therefore being intrinsically modeled in a \emph{metric vector space} that is isomorphic to $\mathcal{R}^3$, so that their typical representation into feature vectors involves their respective coordinates or the coordinates of particularly important, salient points (e.g.~having high curvature). This approach to representing shapes in a geometrical world naturally leads not only to seeking invariance to the above mentioned transformations but also the consideration of implementing pattern comparisons in terms of \emph{distances} between respective feature vectors. The further away two patterns are in the chosen feature space, the less related they are understood to be.

A second mentioned possibility for comparing patterns, namely \emph{statistical joint variation} (e.g.~\cite{JohnsonWichern,Anderson}), consists of modeling the involved features characterizing groups in the dataset as random variables and obtaining estimates for the joint moments, such as the Pearson correlation coefficient. This approach implicitly implements the standardization of the data features, as implied by the own definition of the correlation coefficient. Given two features, the closest this coefficient results to 1, the more intense the tendency of the two random variables being compared to vary together, being therefore interrelated. Contrariwise, the closest the Pearson coefficient results to $-1$, the more intense is the inverse variation between the two variables. Lack of correlation between the two variables is expressed in Pearson coefficient values near to 0.  In the case of comparing two feature vectors, an overall indication of relationship can be estimated in terms of the average of the pairwise correlation between each of the respective features (vector elements).

A third approach to comparing two data elements represented by respective feature vectors consists in \emph{projecting} one of them onto the other, which is often a commutative binary operation. Projections have been typically performed in terms of the scalar product though more general inner products can also be considered. The higher the value of the magnitude of the projection, the more similar the two patterns tend to be. Scalar products are often employed as the initial stage of simple neuronal models (such as McCulloch and Pitts, e.g.~\cite{mcculloch}), frequently involved in convolutive neuronal networks (e.g.~\cite{lecun2015deep,lecun2015deep,pouyanfar2018survey}).  Indeed, the convolution between two vectors can be understood as successive scalar products between one of the vectors (kept fixes) and successively shifted versions of the other vector. The scalar product also provides the basis to the cosine similarity, an approach frequently employed to compare two vectors. However, reflecting its quadratic nature, comparisons by using the scalar product are relatively little sensitive when the two compared patterns are similar, which is often a case of interest in pattern recognition. Comparisons implemented by the scalar product are of \emph{uniform} type, since the projection of one vector onto another scales linearly with their magnitude and are not relative.

A fourth possibility for comparing patterns consists of quantifying the \emph{similarity} between two patterns (e.g.~\cite{costa2022simil}). In practice, this possibility has been approached in terms of comparing the elements of binary sets (i.e.~elements are 0 or 1) by using indices such as Jaccard (e.g.~\cite{Jac:wiki, da2021further}) and S\o rensen-Dice, comparing the elements of more general sets, or by using the cosine similarity index, which corresponds to the smallest angle between the feature vectors respective to the two compared patterns. As observed above, the cosine similarity is also related to the scalar product.  

\begin{figure}
  \centering
     \includegraphics[width=0.7 \textwidth]{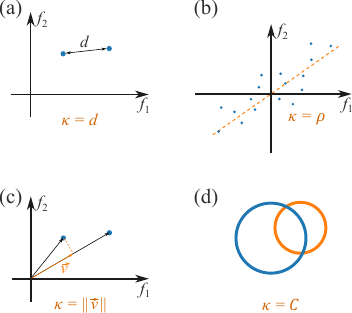}
   \caption{Illustration of the four main approaches to relating data elements as discussed above in terms of a quantified index $\kappa$: (a) Euclidean distance (applicable only in metric spaces); (b) statistical joint variation, such as the Pearson correlation coefficient; (c) projection, as performed by the scalar product typically employed in neuronal networks (applicable only in Hilbert spaces); and (d) Jaccard and coincidence similarity indices, based on the union and intersection of multisets.}\label{fig:approaches}
\end{figure}

Figure~\ref{fig:approaches} summarizes the four types of relationships between data elements as discussed above.

Each of the described approaches have their own advantages and limitations. The following considerations are of special interest while choosing between those possibilities:

I. Several patterns do not leave in a metric space and are not invariant to rotations, the resulting distances are not normalized. In other words, most data are not geometric, so that respective geometric transformations do not direct apply.

II: it is necessary to have a relatively large number of samples so that indices such as Pearson become statistically significant, applicable only between the same features between two groups~\cite{fisher1970statistical,bonett2000sample}. 
In addition, the Pearson correlation coefficient has been found to be unstable in the presence of outliers (e.g.~\cite{kim2015instability}).

III: Approaches obtained by quadratic forms and second order polynomials tend to have limited sensitivity when comparing similar patterns (e.g.~\cite{costa2022simil}). This case includes the scalar product used as initial stage of neurons, Pearson correlation coefficient, and the cosine similarity.

IV: Similarity indices are aimed at quantifying how two data elements (or subclusters) are similar. For instance, we have the cosine similarity, which is directly related to the scalar product between the two respective feature vectors, which implements uniform comparison. Similarity indices based on set and multiset are typically proportional and, therefore, not invariant to translation and rotation, but so are several types of data and patterns. As typically employed, it is mostly restricted to sets or being irrespective to the magnitude of the compared vectors. In the case of Jaccard, it does not take into account the relative interiority between the compared sets. The cosine similarity tends to be little sharp, while not taking into account the magnitudes of the two compared vectors.

Though all the above approaches to relating data elements and subclusters are intrinsically interrelated, the adoption of a specific alternative should be done while taking into account the intrinsic properties of each given problem and dataset.

\section{Uniform and Proportional Feature Spaces}

There are several types of data in science and technology, including \emph{categorical} and \emph{numerical}. The latter can be further subdivided, e.g.~according to the type of numbers (natural, integer, negative, positive, real, etc.) and to the \emph{range} of possible values (e.g.~unbound, limited to an interval $\left[x_{min},x_{max} \right]$, etc.). 

However, there are additional aspects to be considered regarding the values of the features adopted for characterizing a specific dataset, and these have to do with the system from which the data have been obtained. For instance, some specific values or ranges of values can have more or less relevance for the system, in which case they could be accordingly \emph{weighted}.  

Yet another important aspect related to one or more given datasets regards how the data elements should be \emph{compared} to one another. Two situations are of particular interest in the context of the present work \emph{uniform} and \emph{proportional}. By \emph{uniform comparison} it is henceforth understood that the comparison value should not depend on the shared magnitude (i.e.~intersection) of the value of a given feature. For instance, in case we are comparing two real values $a$ and $b$ in terms of a comparison index $\kappa(a,b)$, uniform comparison would require that:
\begin{align}\label{eq:uniform}
   \kappa(\gamma + a, \gamma + b) =  \kappa(a, b),
\end{align}

where $\gamma$ is a generic real value. As an example, we have the \emph{Euclidean distance} between $a$ and $b$ yielding $E(1,2) = E(11,12) = 1$, where $\gamma = 10$, which corresponds to the property of this binary operator $\kappa()$ scaling linearly with $\gamma$, which also implies that:
\begin{align}
   \kappa(\gamma \, a, \gamma \, b) = \gamma \, \kappa(a, b).
\end{align}

As a practical example of a possible application of uniform comparisons, we have a situation in which $a$ and $b$ are weights being transported and one is interested to know the difference of prices between them assuming linear charges are applied. Thus, the cost difference of transporting a weight of $a=2$ k compared to a weight of $b=1$ k will be the same as $a=12$ k and $b=11$ k.

Contrariwise, \emph{proportional comparison} between the values $a$ and $b$ is here understood to be characterized by:
\begin{align}\label{eq:proportional}
   \kappa(\gamma \, a, \gamma \, b) = \kappa(a, b).
\end{align}

This property can be readily achieved by dividing the operator $\kappa(a,b)$ by a quantity $\mathcal{N}(a,b)$ which also scales with $\gamma$, therefore scaling by $\gamma$.

As an example, we may consider the Euclidean distance divided by the average of $a$ and $b$, namely $E(a,b)$, which implies $E(\gamma \, a, \gamma \, b) = \gamma \, E(a, b)$. If we choose as the denominator $\mathcal{N}(a,b)$ corresponding to the arithmetic average between $a$ and $b$, we have that $\mathcal{N}(\gamma \, a, \gamma \, b) =\gamma \, \mathcal{N}(a,b)$.  Thus, we obtain:
\begin{align}
   P(a,b) = \frac{E(a,b)}{\mathcal{N}(a,b)} =  2 \, \frac{ E(a,b) } {a+b},
\end{align}

which, assuming $\gamma=10$, leads to $P[1,2] = P[10,20] = 0.75$.

Observe that, unlike uniform comparisons, the proportional counterpart is necessarily invariant to scaling of the feature values by the same constant $\gamma$. As a consequence, proportional comparisons yield non-dimensional values as a result. In addition, the ratio $a/b$ will also become constant (invariant to scaling).

As an illustration of proportional comparison, we could mention the situation in which one is interested in the relative difference between the lengths of two animals. In this case, it is reasonable to understand that a whale measuring $a=20$ m will be as much longer than another whale with $b=21$ m as an ant with length $0.002$m will be relatively to another ant with $b=0.0021$m.

It is interesting to observe that the choice between adopting uniform and proportional comparison depends more on the purpose of the comparison than on the specific types and values of the data elements.

Given a dataset with $N$ elements, each characterized by $M$ features $x_f$, so that each element $i$ has respective feature vector $\vec{x} = \left[x_1[i], x_2[i], \ldots, x_M[i], \right]$, it is possible to transform it in several manners as a means of implementing normalization of some of the dataset properties. For instance, one could simply take each feature value minus the average of that feature among the whole dataset, which would translate the dataset to its center of mass (each of the new features would necessarily have a null average). As a consequence of this specific normalization, the dataset would become invariant to coordinate translation.

It is interesting to observe that the adoption of uniform or proportional comparison is, on itself, not right or wrong, as the choice depends on the intended purpose of the comparison implied by each specific type of data and respective analysis.

\section{Proportional and Uniform Similarity Indices}

Described recently~\cite{da2021further,costa2022simil,da2022coincidence,costa2023mneurons,da2022brief,costa2021comparing}, the \emph{coincidence similarity index} represents a particularly interesting approach for comparing patterns in terms of their similarity, especially because of its ability to perform strict (sharp) comparison, allowing even the distinction between markedly similar patterns. In addition to being intrinsically normalized between 0 and 1, the coincidence similarity can be applied not only to sets but for the comparison between virtually any pair of mathematical structures. This similarity index has also been found to be robust to the presence of noise and outliers in datasets.  

In addition of taking into account the relative interiority between the compared sets, which is achieved by incorporating the overlap index (e.g.~\cite{vijaymeena2016a}), the potential of application of the coincidence similarity index has been largely enhanced by the adoption of multiset theory (e.g.~\cite{da2021multisets}), allowing the application to real-valued vectors and other mathematical structures. Informally speaking, multisets generalize the traditional concept of set by allowing the elements to appear repeatedly, which is quantified in terms of the \emph{multiplicity} of each of the constituent elements. Multiset theory allows the set operations of union and intersection, which are required for the calculation of the Jaccard similarity index, to be replaced by the operations of maximum and minimum, respectively. Introduced more recently, the description of a general approach to allowing multiset operations to be performed between vectors with possibly negative entries has extended even further the potential of extensions and applications of the coincidence similarity index.  Thanks to its several interesting properties and despite its recent introduction, the coincidence similarity index has already been successfully applied to a wide range of applications (e.g.~\cite{benatti2022neuromorphic, domingues2022identification, tokuda2023cross,benatti2023quantifying,benatti2024simple,benatti2023multilayer, benatti2023two}).

The similarity indices considered in the present work are based on concepts and properties of \emph{multisets} (e.g.~\cite{da2021multisets}) which, basically, can be understood as a generalization of the more traditional concept of \emph{set} in which elements are allowed to appear several times. The number of times each element appears is taken as its respective \emph{multiplicity}.  For instance, consider the following example of a multiset:
\begin{align}
  R = \{\!|\ a, a, a, b, d, d, d, d |\!\}  = [\![ \left( a, 3\right); \left(b, 1 \right); \left(d, 4 \right) ]\!]  \nonumber.
\end{align}
We have that element `a' has multiplicity 3, element `b' has multiplicity 1, and element `d' has multiplicity 4. Observe that the fact that an element $x_i$ has multiplicity $a$ in a given multiset $R$ can be expressed as $ \left( x_i, m^A_i = a \right)$, with $i = 1, 2, \ldots, N$, where $N$ is the total number of elements in the multiset.

The \emph{cardinality} of a multiset $R$ can be defined as:
\begin{align}
  |R| = \sum_{i=1}^N m^R_i.
\end{align}

For instance, in the case of $R$ above, we would have $|R| = 8$.
As with sets, it is possible to perform operations of \emph{union} and \emph{intersection} between multisets.  For instance, consider a second example of multiset as:
\begin{align}
  S = \{\!|\ a, a, b, b, b, d |\!\}  = [\![ \left( a, 2\right); \left(b, 2 \right); \left(d, 1 \right) ]\!]  \nonumber
\end{align}

The \emph{union} between $A$ and $B$ can be readily obtained by considering the maximum value between the respective multiplicities in both multisets.  In the case of the two multisets above, we would have:
\begin{align}
  R \cup S = [\![ \left( a, 3\right); \left(b, 2 \right); \left(d, 4 \right) ]\!] =
    \{\!|\ a, a, a, b, b, d, d, d, d |\!\}   \nonumber.
\end{align}
The \emph{intersection} between two multisets can be similarly obtained by taking the minimum, instead of the maximum, between the multiplicities. Thus, we have that:
\begin{align}
  R \cap S = [\![ \left( a, 2\right); \left(b, 1 \right); \left(d, 1 \right) ]\!] =
    \{\!|\ a, b, d |\!\}   \nonumber
\end{align}

It is possible to extend multisets to have real multiplicity (e.g.~\cite{costa2023mulsetions}), which allows real-valued vectors to be represented a multisets. For instance, we could have:
\begin{align}
  &A = [\![ \left( -0.5, 1.3\right); \left(1.0, -2.2 \right); \left(-1.2,  -3\right) ]\!]  \nonumber  \\
  &B = [\![ \left( -0.5, 2.1\right); \left(1.0, 2.0 \right); \left(-1.2,  -2\right) ]\!]  \nonumber  
\end{align}

The \emph{cardinality} of a real-valued multiset $A$ can be defined as:
\begin{align}
  |A| = \sum_{i=1}^N |m^A_i|
\end{align}

The union and intersection between two real-values multisets can be performed as described in~\cite{costa2023mulsetions}, which involves representing separately the positive and negative multiplicities in each multiset.  In the case of $A$ and $B$ above, we would have:
\begin{align}
  &A^P = [\![ \left( -0.5, 1.3\right); \left(1.0, 0 \right); \left(-1.2,  0\right) ]\!]  \nonumber  \\
  &A^N = [\![ \left( -0.5, 0\right); \left(1.0, -2.2 \right); \left(-1.2,  -3\right) ]\!]  \nonumber  \\
  &B^P = [\![ \left( -0.5, 2.1\right); \left(1.0, 2.0 \right); \left(-1.2,  0\right) ]\!]  \nonumber  \\
  &B^N = [\![ \left( -0.5, 0\right); \left(1.0, 0 \right); \left(-1.2,  -2\right) ]\!]  \nonumber  
\end{align}

The multiset \emph{union} between these two sets can now be obtained as:
\begin{align}
  A \cup B = [\![ \left( a, 3\right); \left(b, 2 \right); \left(d, 4 \right) ]\!]   \nonumber
\end{align}

Similarly, the \emph{intersection} between those two real-valued multisets can be calculated as:
\begin{align}
  A \cap B = [\![ \left( a, 3\right); \left(b, 2 \right); \left(d, 4 \right) ]\!]   \nonumber
\end{align}

Known and having been applied for a long time (e.g.~\cite{da2021further}), the Jaccard similarity index between two non-empty sets $A$ and $B$ can be defined as:
\begin{align}
    \mathcal{J}(A,B) = \frac{A \cap B}{A \cup B}.
\end{align}

The relative interiority (also overlap, e.g.~\cite{vijaymeena2016a}) between the two compared sets can be expressed as:
\begin{align}
    \mathcal{I}(A,B) = \frac{A \cap B}{\min\left(|A|,|B| \right)}  
\end{align}
where $||$ indicates the cardinality of the set.

As discussed in~\cite{da2021further}, the Jaccard index does not take into account the interiority between the compared sets, therefore providing the same similarity value in those cases.  Motivated by this issue, the \emph{coincidence similarity index} has been introduced, combining the Jaccard and interiority indices as follows:
\begin{align}
    \mathcal{C}(A,B) = \left[ \mathcal{J}(A,B) \right]^D \, \mathcal{I}(A,B),
\end{align}
where $D$ is a positive real value controlling how strict the comparison performed by the Jaccard index. The higher the value of $D$, the more strict the comparison is.

The Jaccard, interiority, and coincidence complex indices for comparing two non-zero vectors (multisets) $\vec{m}_A$ and $\vec{m}_B$ having possibly non-negative values~\cite{costa2023mulsetions} can be expressed as:

\begin{align}
    &\mathcal{J}(\vec{m}_A,\vec{m}_B) = \frac{\sum_i \{\min(m^P_{i,A}, m^P_{i,B})  + \min(|m^N_{i,A}|, |m^N_{i,B}|)\}}{\sum_i \{\max(m^P_{i,A}, m^P_{i,B}) + \max(|m^N_{i,A}|, |m^N_{i,B}|)\}}, \label{eq:Jaccard} \\
    &\mathcal{I}(\vec{m}_A,\vec{m}_B) = \frac{\sum_i \{\min(m^P_{i,A}, m^P_{i,B})  + \min(|m^N_{i,A}|, |m^N_{i,B}|)\}}{\min \{ \sum_i|m_{i,A}|, \sum_i|m_{i,B}| \}  },  \label{eq:Inter}  \\
    &\mathcal{C}(\vec{m}_A,\vec{m}_B) = \left[ \mathcal{J}(\vec{m}_A,\vec{m}_B) \right]^D \, \mathcal{I}(\vec{m}_A,\vec{m}_B),
\end{align}

where the superscripts $N$ and $P$ indicates the positive and negative portions of the original vectors.  Observe that both the numerator and denominator in Equations~\ref{eq:Jaccard} and~\ref{eq:Inter} are non-negative values, hence neither these indices (as well as the coincidence similarity) can take negative values.

It can be verified that the Jaccard, interiority and coincidence similarity indices implement proportional comparisons between the two given non-zero vectors or multisets.  That is a consequence of the involved ratio between two quantities with the same unit, which also leads to non-dimensional values normalized in the interval $[0,1]$.

Because the coincidence similarity implements a particularly strict comparison near the origin of the coordinate system, clusters in uniform spaces which are near this position may be broken into subclusters.  In order to address this tendency, and also to provide some regularization at points near the origin of the coordinate system, a small constant $\delta$ can be added at both the numerator and denominator of Equation~\ref{eq:Jaccard}.  

In this work, we also define the uniform similarity index ($\mathcal{U}$) which quantifies the similarity between two variables according to the uniform comparison presented in Equation~\ref{eq:uniform}. The uniform similarity index can be expressed as:
\begin{align}
    \mathcal{U}(\vec{m}_A,\vec{m}_B) = \frac{1}{1+\sum_i |m_{i,A} - m_{i,B}|} 
\end{align}

\begin{figure}
  \centering
     \includegraphics[width=0.7 \textwidth]{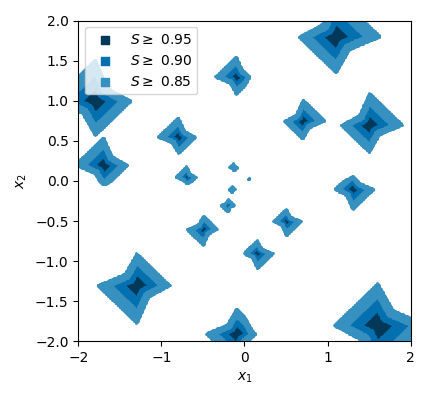}
 \caption{The \emph{receptive fields} characterizing the coincidence similarity index for proportional comparisons. Each field is composed of the regions of comparison values indicated by colors according to the legend. As a characteristic of proportional comparison, the size of the receptive fields increases linearly with the distance from the origin of the coordinate system. The implemented comparisons are particularly strict near the origin of the feature space, with a singularity at that origin. }\label{fig:restrictivity}
\end{figure}

\begin{figure}
  \centering
     \includegraphics[width=0.7 \textwidth]{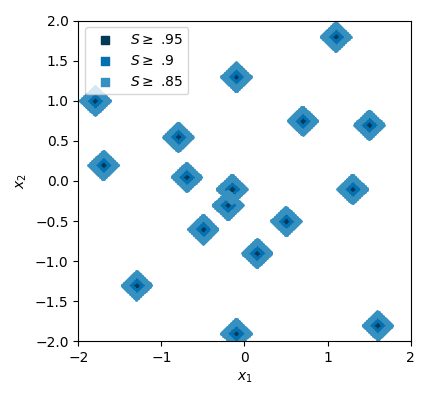}
 \caption{The \emph{receptive fields} characterizing the uniform similarity index for performing uniform comparisons. Each field is composed of the regions of comparison values indicated by colors according to the legend. As a characteristic of proportional comparison, the size of the receptive fields remains constant with the distance from the origin of the coordinate system. The implemented comparisons are equally selective throughout the feature space. }\label{fig:restrictivity2}
\end{figure}

Figure~\ref{fig:restrictivity} and~\ref{fig:restrictivity2} illustrate (in blue) some of the highest level-sets defined by the coincidence similarity and the uniform similarity respectively to sets of points in a two-dimensional feature space.  For simplicity's sake, each of the obtained classification regions will be henceforth called a \emph{receptive field}. Several interesting aspects can be identified from these two figures.  First, we have that the size of the receptive fields implied by the coincidence similarity increases linearly with the respective distance to the origin of the coordinate center.  This is verified in proportional comparison approaches.  On the other hand, the receptive fields defined by the uniform similarity all have the same size, which is typical of uniform feature spaces. The shapes of the receptive fields obtained for the coincidence and uniform similarity are distinct because of the non-linear nature of the former index, which involves both the Jaccard and interiority indices.

\begin{figure}
  \centering
     \includegraphics[width=0.7 \textwidth]{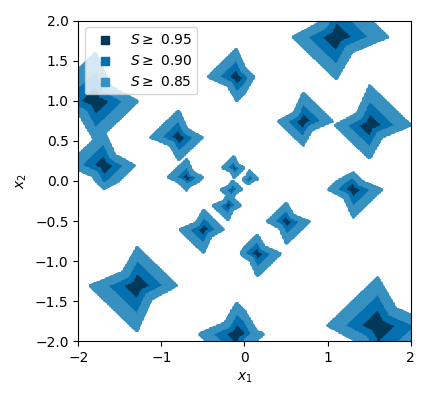}
 \caption{The coincidence similarity receptive fields from Fig.~\ref{fig:restrictivity} regularized by using $\delta=0.5$. The origin of the coordinate axis is no longer associated to a singularity.}\label{fig:restrictivity1}
\end{figure}

Figure~\ref{fig:restrictivity1} illustrates the effect of this type of regularization on the coincidence similarity receptive fields (Fig.~\ref{fig:restrictivity}) respectively to $\delta=0.5$. Though a relatively large value of $\delta$ has been used in this example for the sake of enhanced visualization, the remainder of the present work adopts $\delta=0.01$.

\section{Agglomerative Clustering and Dendrograms}

A dendrogram is a graphical representation that resembles a tree and is useful for representing hierarchical relationships between different sets of data. In general, a dendrogram is a map that shows how different groups of data are connected to each other. This tool is widely used in statistical analysis, especially in hierarchical clustering methods, where each branch of the dendrogram represents a possible division between the groups, indicating a level of similarity or difference between them.

Henceforth, it is assumed that the dendrograms have two main axis.  The horizontal axis indicates to the labels of the involved data elements, which can be placed in arbitrary order while avoiding crossings between the respective dendrogram branches.  The vertical axis, with values increasing from the bottom to the top of the dendrogram, corresponds to some measurement of the \emph{lack of relationship} (e.g.~distance or dissimilarity) between the data elements, progressing from low values (larger relationship) toward higher values (smaller relationship).  In the case of measurements normalized in the interval $[0,1]$, as is the case with the uniform and proportional multiset similarity indices , $C$ and $U$ considered in the present work, the vertical axis can be defined as corresponding to the quantities $1-C$ and $1-U$, respectively.

The applications of a dendrogram are numerous and span various domains of knowledge. In biology, for instance, it is employed to illustrate the evolutionary relationships between species. In psychology, it can assist in comprehending the structure of personality traits. Among the benefits of employing a dendrogram is its capacity to visually convey the intricacy of data in an intuitive manner.

The Ward's method~\cite{ward1963hierarchical,duda2000pattern} represents one of the most well-known and frequently applied agglomerative clustering methods. This approach progressively combines the clusters in order to minimize the variance within the clusters. At each stage of the process, the total sum of the variances within each cluster is calculated, and the existing clusters are merged in a way that will result in the smallest possible increase in this dispersion measure. In contrast to the conventional approach of merely considering the shortest distance between data points or clusters, the Ward's method evaluates the impact of each merger on the internal homogeneity of the clusters.  However, this approach has a tendency of finding clustering structures even when they are weak or non-existent (e.g.~\cite{tokuda2022}).

The length of the stem of a specific cluster in a given dendrogram can be related to the respective support for that cluster, in the sense that the longer the stem, the more likely the existence of a real cluster.  However, this is not always the case, as the heights of the branches can be relatively long even in the case of uniformly random noise (e.g.~\cite{tokuda2022}).  In particular, the Ward's approach has a tendency to highlight clustering, providing indication for clusters even in the case of statistical fluctuations.

\begin{figure}
  \centering
     \includegraphics[width=0.7 \textwidth]{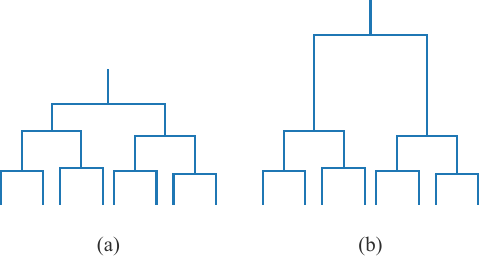}
   \caption{Examples of dendrograms indicating no presence of clustering structure (a) and presence of two main clusters (b), revealed by the substantially longer lengths of the respective stems.}\label{fig:2dendrogram}
\end{figure}

In the case of a uniformly random set of data elements, it would be expected that all stems would have similar Figure~\ref{fig:2dendrogram}(a) illustrates this situation. The lengths of all stems are about equal, indicating uniform distribution of the original data elements and respective sub-cluster. On the other hand, the dendrogram illustrated in Figure~\ref{fig:2dendrogram}(b) provides a strong indication about the presence of two main branches, reflected in the substantially longer lengths of the respective stems.

Furthermore, the numerical values of the lengths can be employed to define an objective measure that indicates the possibility of groups in the original data, represented as a respective dendrogram.

Let us represent the set of branch lengths as $L$, where $L = {l_1, l_2, \dots, l_n}$. Here, $l_i$ corresponds to a given branch length in the dendrogram. The standard deviation of $L$ can be estimated as:
\begin{equation}
    \sigma_L = \sqrt{\frac{\sum_i (l_i-\overline{L})^2}{n}}
\end{equation}
where $\overline{L}$ is the average value of $L$.

An index $H$ quantifying the overall \emph{homogeneity} of the existing groups can then be expressed as:
\begin{equation}
    H = \frac{\overline{L}}{\sigma_L}
\end{equation}

Observe that $H$ is non-dimensional, being therefore invariant to the overall magnitude or scale of the original measurement of data relationship. The larger the value of $H$, the more uniform the original data can be understood to be, therefore providing little indication about the presence of possible clusters.

It is also possible to consider an index $r_i$ providing indication that a specific branch $i$ corresponds to a cluster:
Relevancy
\begin{equation}
    r_i = \frac{l_i}{\overline{L}}
\end{equation}

Similarly to the index $H$, the above quantity is also non-dimensional. Data groups associated to a large value of $r_i$ can be understood to as possible clusters.

\begin{figure}
  \centering
     \includegraphics[width=1. \textwidth]{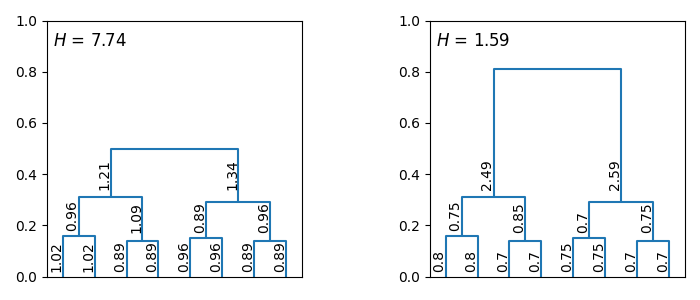} \\
     \hspace{.5cm} (a) \hspace{6.5cm} (b)
   \caption{The values of the indices $H$ and $r_i$ obtained for two distinct dendrograms corresponding to a mostly uniform data (a) and a data set subdivided into two main groups (b). 
 The vertical axis corresponds to some measurement quantifying lack of relationship (e.g.~distance, joint-variation, similarity) between the involved groups, so that more closely related pairs of groups merge sooner along the respective axis.}\label{fig:2dendrogram_meansures}
\end{figure}

Figure~\ref{fig:2dendrogram_meansures} illustrates the abode described indices $H$ and $r_i$ respectively to two distinct dendrograms.  

In the first case, shown in Figure~\ref{fig:2dendrogram_meansures}(a), which is similar to that shown in Figure~\ref{fig:2dendrogram_meansures}, the dendrogram indicates a more uniform relationship between the sub-clusters at successive hierarchical levels. On the other hand, the dendrogram shown in Figure~\ref{fig:2dendrogram_meansures}(b) provides an illustration of a more heterogeneous data structure characterized by a division into two main groups.   As could be expected, the value of $H$ obtained in the former case is significantly larger than that obtained for the second dendrogram, suggesting that the latter situation is more likely to present a clustering structure.  

Also shown in Figure~\ref{fig:2dendrogram_meansures} are the individual relevance indices $r_i$ associated to each possible sub-cluster in the original respective data.  Similar results have been obtained among all branches with and across the two dendrograms, with the exception of the two highest branches in Figure~\ref{fig:2dendrogram_meansures}(b), whose substantially larger values indicate the present of a possible respective cluster in the original dataset.

\section{Clusters in Uniform and Proportional Spaces}

Developments related to recognizing patterns can benefit from a better understanding of how clusters, or groups, of data elements are expected to appear. In other words, it is interesting to have some possible \emph{models} of how clusters (e.g.~\cite{tokuda2022}) appear in uniform and proportional spaces. This interesting subject is considered in the present section. Henceforth, we understand that the elements of a given data set are characterized in terms of respective features or measurements, which can be represented as points in a corresponding \emph{feature space}.

One important initial aspect concerns the own \emph{definition} of what a cluster is. While no definitive or consensual formal answer to this question seems to exist, a relatively informal approach has been frequently considered in pattern recognition which, given a set of data elements, understands that:

\begin{center}
\begin{tcolorbox}[colback=white, boxrule=-1pt, width=\textwidth-4cm]
\emph{A cluster corresponds to a subset of elements which are more \textbf{related} one another than with the remainder of the elements in the given dataset.}
\end{tcolorbox}
\end{center}

One limitation with the above concept is that there are several manners in which a set of points can relate one another, which include but are not limited to \emph{distance}, \emph{similarity}, or \emph{joint variation}. In addition, the points within a cluster can be separated one another in a virtually infinite number of manners, including uniform or normal spatial distributions. Several types of cluster borders or shapes can also exist.

\begin{figure}
  \centering
     \includegraphics[width=0.5 \textwidth]{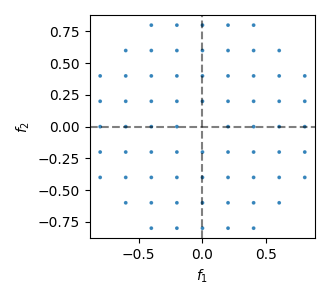}\\
     \hspace{1.5cm}(a)\\ \vspace{.5cm}
     \includegraphics[width=1 \textwidth]{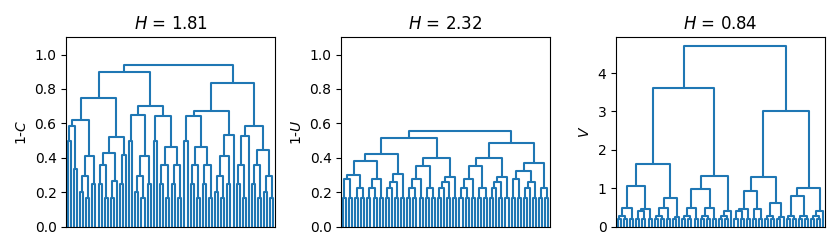}\\
     \hspace{.8cm} (b) \hspace{3.8cm} (c) \hspace{3.8cm} (d)\\
   \caption{Dendrograms obtained respectively to a model cluster in a \emph{uniform} space (a) by the coincidence similarity (b), uniform similarity (c), and Ward's (d) methods.  The respective value of the index $H$ are also provided. The coincidence similarity led to the most well-balanced dendrogram providing a little indication of the presence of subclustering structures, while the Ward's methodology led to unbalanced dendrogram which also strongly indicates subcluster structures.}\label{fig:uniforme0_comparison}
\end{figure}

In the present work, we focus on clusters with points which are uniformly distributed within a delimiting region. More specifically, the structure illustrated in Figure~\ref{fig:uniforme0_comparison}(a), respectively to a two-dimensional \emph{uniform feature space}, or simply \emph{uniform space}, is henceforth adopted as a basic model of a cluster in a uniform space. This figure also presents the dendrograms obtained respectively to the coincidence similarity (b), uniform similarity (c), and Ward's (d)methodology.

As could be expected, the first merging of the points takes place at the same dendrogram height $1-U$. This is not necessarily the case for the two other hierarchical clustering approaches. In addition, the relatively low maximum height of the dendrogram obtained for uniform similarity indicates more effectively that the set of points is uniform. At the same time, the relative lengths of the stems are short, providing a little indication of subclustering structures, which is indeed the case given that the original set corresponds to an orthogonal lattice. Contrariwise, the two long stems at the top of the Ward's dendrogram provide a substantial indication of the presence of two clusters, which is not the case. The dendrogram obtained for the coincidence similarity is similar to that obtained by the uniform similarity, though presenting a larger height.

Interestingly, the obtained values of $H$ confirm the above considerations, indicating that the most uniform dendrogram is that obtained by the uniform clustering approach (Fig.~\ref{fig:uniforme0_comparison}b), followed by the results obtained respectively to the proportional and Ward's methods.  In particular, the latter approach can be understood to strongly suggest the presence of heterogeneity and possible clusters in the original data, which is not the case for this particular data set.

More specifically, we have a set of points distributed as an orthogonal lattice comprised within a circular boundary. The important property of this model is that the nearest distance between each point and its neighbors is equal for any point in the cluster.

\begin{figure}
  \centering
     \includegraphics[width=0.5 \textwidth]{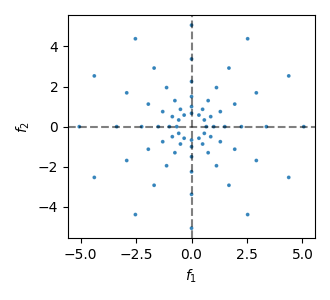}\\
     \hspace{1.cm}(a)\\ \vspace{.5cm}
     \includegraphics[width=1 \textwidth]{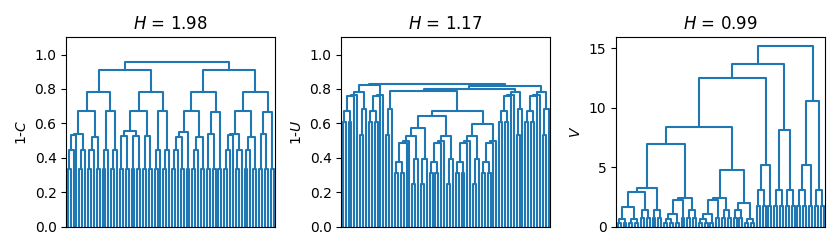}\\
     \hspace{.8cm} (b) \hspace{3.8cm} (c) \hspace{3.8cm} (d)\\   
   \caption{Dendrograms obtained respectively to a model cluster in a \emph{proportional} space (a) by the coincidence similarity (b), uniform similarity (c), and Ward's  (d) methods.  The values of $H$ are also respectively provided. The coincidence similarity led to the most well-balanced dendrogram providing little indication of the presence of subclustering structures, while the Ward's methodology led to an unbalanced dendrogram strongly indicating subclustering structures. 
 }\label{fig:proportional0_comparison}
\end{figure}

Figure~\ref{fig:proportional0_comparison}(a) presents the model of cluster adopted in the present work in the case of \emph{proportional spaces}, also including respective dendrograms obtained considering coincidence similarity (b), uniform similarity (c), and Ward's methodology (d).

As can be observed in this figure, the dendrogram obtained for the coincidence similarity has the points first merging at the same height $1-C$, which is not the case for the other two considered hierarchical clustering methods.  This important characteristic of the original data is properly indicated in the dendrogram obtained for the coincidence similarity, which also has relatively short stems at all hierarchical levels, therefore providing a little indication of subclustering structures, which is indeed the case given the proportional regularity of the original data. This is also verified for the dendrogram obtained by using the uniform similarity, though the heights of the first mergings vary intensely, indicating a non-uniformity that is not present in the original data. The Ward's dendrogram not only indicates a strong indication of the presence of heterogeneous data distribution possibly involving subclustering structures (long stems at the higher hierarchies), which is not the case given the proportional regularity of the original data.

Again, the obtained values of $H$ provide an objective quantification of the overall indication of homogeneity provided by each of the three considered approaches, with the result obtained by the proportional clustering corresponding to the most homogeneous dendrogram structure, which is indeed the case for this particular dataset.  The Ward's approach again suggests heterogeneous overall subclustering structure.

By comparing the structures in Figures~\ref{fig:uniforme0_comparison}(a) and~\ref{fig:proportional0_comparison}(a) with the receptive fields in Figures~\ref{fig:restrictivity} and~\ref{fig:restrictivity2}, the intrinsic congruence between the respective densities of points and size of receptive fields can be readily appreciated.  This congruence reflects the intrinsic suitability between uniform and proportional data types and respective comparison approaches.

All in all, we have that the uniform similarity approach provides a particularly suitable representation of the relationship between the points for the adopted cluster model in uniform spaces, while the coincidence similarity method provides a more effective representation of the relationships in the case of the proportional cluster model. The Ward's approach tended, in both situations, to provide indication about the presence of subclustering structures, which is not the case for either uniform or proportional cluster models.

\section{Methodology}\label{sec:concepts}

The construction of dendrograms represents a fundamental resource to be considered in hierarchical cluster analysis. In the case of the present work, this begins with the calculation of a similarity matrix, which quantifies the similarity between every pair of elements. The resulting matrix is then used to determine which clusters are to be combined at each hierarchical stage.

In the initial stage, each element is regarded as an independent cluster.  Subsequently, an agglomerative approach is employed, whereby the two most analogous clusters are merged to form a new cluster. This process is repeated until all elements have been incorporated into a single cluster. At this point, the similarity between each new cluster is determined as the arithmetic average of the similarity between its elements. Though, in analogy to alternative traditional agglomerative clustering methods such as single and complete linkage (e.g.~\cite{duda2000pattern}), it is also possible to obtain similarity-based hierarchical clustering approaches considering the minimum and maximum similarity between subclusters.

The dendrogram is often obtained as a means to visualize the results yielded by agglomerative clustering, allowing the succession of mergings to be represented in an effective hierarchical manner. In particular, the height of the lines connecting the clusters reflects the dissimilarity (1-$S$) between them. The lower the height at which a merging takes place, the greater the similarity between the clusters.

In order to assess and compare the efficacy of the considered clustering algorithms, we considered a metric of accuracy reflecting the correct identification of the clusters. 

\begin{figure}
  \centering
     \includegraphics[width=0.5 \textwidth]{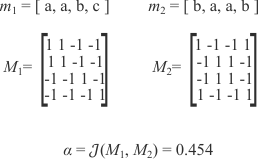} 
   \caption{Illustration of the index $\alpha$ adopted as a means to quantify the performance of the hierarchical clustering.  Given a vector $m_1$ containing the original categories, as well as the vector $m_2$ of the categories assigned by the clustering algorithm, respective matrices $M_1$ and $M_2$ are obtained reflecting respective joint membership between all original elements. 
   The Jaccard index is then calculated according to Eq.~\ref{eq:Jaccard}, resulting the value $\alpha = 0.454$ in the case of this specific example.}\label{fig:acc_ex}
\end{figure}

The employed performance evaluation index was defined as the proportion of correct predictions made by the model in relation to the total number of predictions. More specifically, it corresponds to the Jaccard similarity index between two matrices defined from the original and obtained groups, as illustrated in Figure~\ref{fig:acc_ex}. Therefore, this index necessarily takes real values ranging from 0 to 1.

This metric was selected because it is not contingent on the order of the groups, or the label assigned to each one. Consequently, this measure of accuracy considers the  congruence between the structure of the original data and the algorithm's capacity to effectively capture that clkustering structure.

\section{Results and Discussion}\label{sec:results}

This section presents the experimental results aimed at better understanding the relative characteristics and performance of the agglomerative methods based on uniform and proportional similarity, as well as the traditional Ward's approach. First, a preliminary experiments are reported respectively to the three clustering approaches applied to two clusters in uniform feature space, a single cluster in a proportional space, and two clusters in a proportional feature space.  Subsequently, a more systematic comparison between the three considering clustering methods are described and discussed respectively to three types of clusters in a uniform feature space.  The Jaccard similarity is henceforth employed with a regularization $\delta=0.1$.  Unless otherwise indicated, the coincidence similarity is henceforth applied with $D=1$.

\subsection{Preliminary Comparison}\label{subsec:preliminary}

\begin{figure}
  \centering
     \includegraphics[width=1 \textwidth]{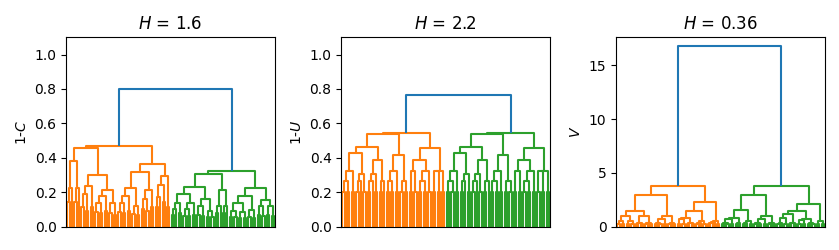}\\
     \hspace{.5cm} (a) \hspace{3.5cm} (b) \hspace{3.8cm} (c)\\ \vspace{.5cm}
     \includegraphics[width=1 \textwidth]{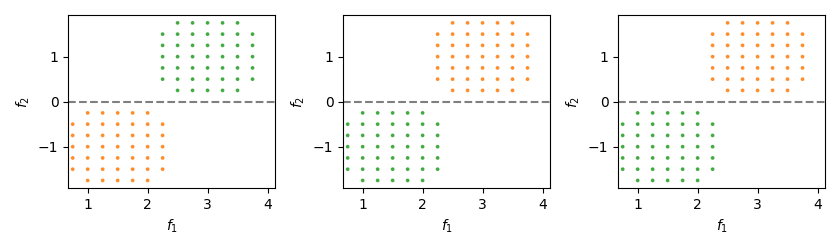}\\
     \hspace{.5cm} (d) \hspace{3.5cm} (e) \hspace{3.8cm} (f)
 \caption{Hierarchical clustering of a dataset containing two separated clusters in a \emph{uniform} space as performed by coincidence similarity (a), uniform similarity (b), and Ward's method (c). The two clusters identified by each of these methodologies are shown in (d), (e), and (f), respectively. Though all three approaches were capable of properly identifying the two original clusters, the uniform similarity can be understood as resulting particularly effective, followed by the coincidence similarity and Ward's methods (see text for additional discussion). 
 }\label{fig:uniforme_comparison}
\end{figure}

A first relevant situation to consider is presented in Figure~\ref{fig:uniforme_comparison}, involving two circular clusters separated along the diagonal direction. The figure shows the dendrograms respectively obtained by the agglomerative methods based on the proportional (a) and uniform (b) similarity methods, as well as by the Ward's methodology (c). The respective results are presented in Figures~\ref{fig:uniforme_comparison}(d-f).

Despite being near one another, the two clusters were properly identified by all three considered methodologies.  However, the approach based on the uniform similarity yielded two well-balanced main branches in the respectively obtained dendrogram shown in Figure~\ref{fig:uniforme_comparison}(b), being also characterized by a relatively high value of $H$. In addition, in that case, the length of the stems increase in less abrupt manner than those in the dendrogram obtained by the Ward's methodology (c). It is also worth noticing that the Ward's approach is based on minimizing data variance, being therefore intrinsically adapted to clusters which have the elements distributed according (even if approximately) to the normal density, including circularly symmetric an elongated data elements distributions.  At the same time, the proportional similarity agglomerative clustering led to two moderately balanced main branches in the respective dendrogram (c).  

In summary, all the three approaches were capable of properly identifying the two original clusters, but the uniform similarity methods led to a well-balanced dendrogram emphasizing only the two main branches. Despite being more specific for proportional feature spaces, the proportional similarity approach was still able to provide an interesting result characterized by moderately balanced hierarchies and little indication of subclustering structures.

\begin{figure}
  \centering
     \includegraphics[width=1 \textwidth]{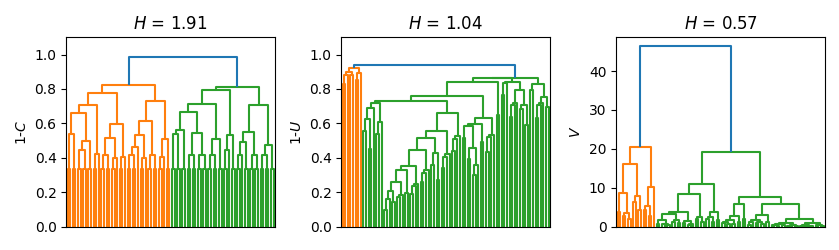}\\
     \hspace{.5cm} (a) \hspace{3.5cm} (b) \hspace{3.8cm} (c)\\ \vspace{.5cm}
     \includegraphics[width=1 \textwidth]{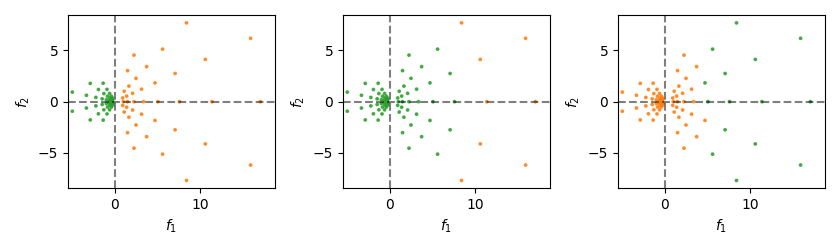}\\
     \hspace{.5cm} (d) \hspace{3.5cm} (e) \hspace{3.8cm} (f)
 \caption{Hierarchical clustering of a dataset containing two separated clusters in a \emph{proportional} space as performed by coincidence similarity (a), uniform similarity (b), and Ward's method (c). The two clusters identified by each of these methodologies are shown in (d), (e), and (f), respectively. Only the coincidence similarity method allowed proper separation of the two original groups,  also characterized by well-balanced branches in the respective dendrogram (see text for additional discussion). 
 }\label{fig:proportional_comparison}
\end{figure}

Figure~\ref{fig:proportional_comparison} refers to the important situation in which two clusters are present in a proportional feature space, which are close one another while being horizontally separated by the $y-$axis (see Figs.~\ref{fig:proportional_comparison}d-f). Neither the uniform similarity nor the Ward's approaches were able to properly identify the two original clusters. In addition, completely unbalanced dendrograms have been obtained by those two methods.  As shown in Figures~\ref{fig:proportional_comparison}(a-b), not only the two original clusters have been properly identified, but the respective dendrogram resulted well-balanced and having the length of the stems corresponding to the two original clusters longer than those at lower clustering hierarchies.  The obtained values of $H$ tend to corroborate these observations.

\begin{figure}
  \centering
     \includegraphics[width=1 \textwidth]{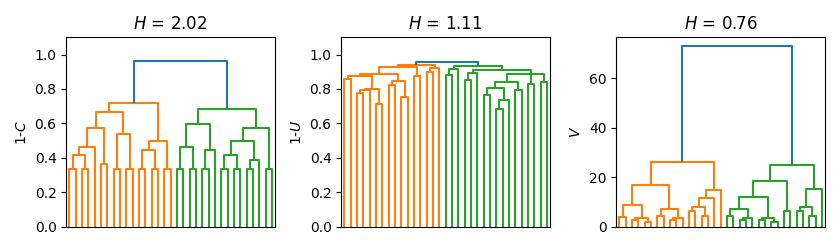}\\
     \hspace{.5cm} (a) \hspace{3.5cm} (b) \hspace{3.8cm} (c)\\ \vspace{.5cm}
     \includegraphics[width=1 \textwidth]{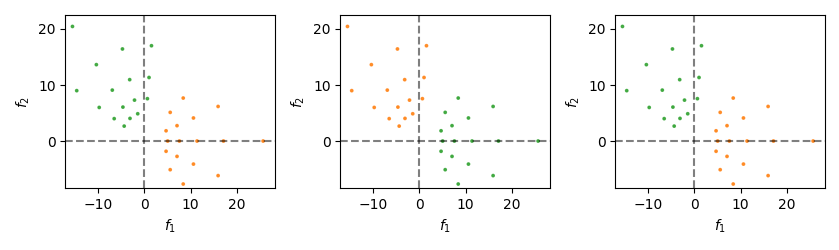}\\
     \hspace{.5cm} (d) \hspace{3.5cm} (e) \hspace{3.8cm} (f)
 \caption{Hierarchical clustering of a dataset containing two separated clusters at the same distance from the origin in a \emph{proportional} space as performed by coincidence similarity (a), uniform similarity (b), and Ward's method (c). The two clusters identified by each of these methodologies are shown in (d), (e), and (f), respectively.  Though all three methods could identify the two original clusters, the results obtained in the case of the coincidence similarity can be understood as being more suitable in the sense of involving a well-balanced dendrogram with respective stem lengths that are less pronounced than those obtained by the Ward's approach (see text for additional discussion). }\label{fig:proportional2_comparison}
\end{figure}

Figure~\ref{fig:proportional2_comparison} presents the dendrograms and clustering results respectively to two original clusters in a proportional feature space which are separated angularly. Though the three considered methods have been able to identify the two original clusters, the uniform similarity approach yielded a respective dendrogram which is not only unbalanced (with relatively large value of $H$), but also presents little evidence for the presence of two main clusters, as indicated by the markedly short lengths of the respective two main stems (Fig.~\ref{fig:proportional2_comparison}b). The Ward's approach led not only to an unbalanced dendrogram but also provided indication of subclustering structures within the two main identified clusters, which is not the case for this dataset. The proportional similarity approach yielded a dendrogram (Fig.~\ref{fig:proportional2_comparison}a) which is moderately balanced at the highest hierarchical level (the main two clusters) and mostly well-balanced structure along the lower levels, with little indication of subclustering structures.  Several of the above discussed results have been substantiated by the respectively obtained values of $H$.

All in all, the reported experiments and results suggest that, at least for the considered clustering structures and parameter configurations, the uniform similarity method is particularly suitable for identifying clusters in uniform features spaces. At the same time, though the proportional similarity approach resulted particularly effective to identify clusters in proportional features spaces, it was also capable of providing interesting results in the case of uniform feature space, identifying properly the two clusters in all considered situations. The Ward's methodology, though able to properly identify the clusters in two of those configurations, tended to provide strong indication of subclustering structures in all cases.

\subsection{Comparing Hierarchical Clustering Approaches}

\begin{figure}
  \centering
     \includegraphics[width=.7 \textwidth]{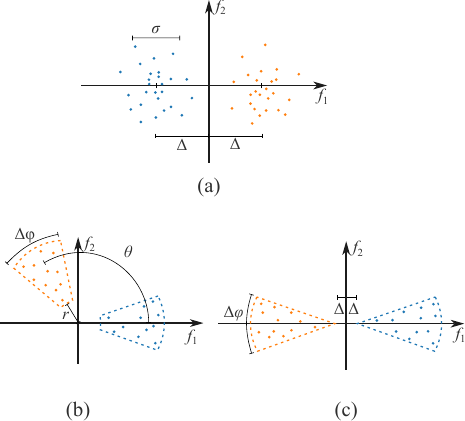}
 \caption{The three types of two-dimensional cluster structures considered in the present work: (a) two normal densities with average over the $x-$axis and with distance $\Delta$ to the origin of the coordinate system; (b) two clusters consisting of points uniformly distributed within a section of a disk, with relative angular separation of $\theta$; (c) the same as before, but with $r=0$, fixed angular position, and each with distance $\Delta$ to the origin of the coordinate system.}\label{fig:cluster}
\end{figure}

In order to obtain experimental indication about the relative performance of the three main approaches considered in the present work --- namely uniform, proportional, and Ward's, these methods have been applied to several instances of the three basic cluster structures illustrated in Figure~\ref{fig:cluster}, all of them involving \emph{uniform} 2D feature spaces. Proportional feature spaces have not been considered in this section since the preliminary results described in Section~\ref{subsec:preliminary} indicated that clusters in proportional spaces seem to be only properly characterized by the proportional similarity methodology.

The situation depicted in Figure~\ref{fig:cluster}(a) consists of $n=50$ points distributed uniformly within each of two circular clusters symmetrically placed with respect to the origin of the coordinate system. The distance $\Delta$ from the origin is varied as $\Delta=0.1, 0.105, \ldots,0.5$.

The two situations shown in Figures~\ref{fig:cluster}(b) and (c) involves two clusters corresponding to points distributed within circular sections. In the former case, these clusters are rotated by $\theta = 10, 14, \ldots, 90$ degrees, while keeping $\Delta \phi = 10$ degrees and $r=0.1$. In the latter case, the two clusters are organized as shown in Figure~\ref{fig:cluster}(c), with $\Delta=0, 0.01, \ldots,0.2$.

A total of $T=100$ samples of each of the three types of clustering structures described above are obtained and fed to each of the three methods, with the respective performance being quantified by the accuracy index $\alpha$. These experiments adopt $\sigma=0.1$, $\Delta \phi = 10$ degrees, and $r=0.1$. The present section describes and discussed the respectively obtained results.

\begin{figure}
  \centering
     \includegraphics[width=1 \textwidth]{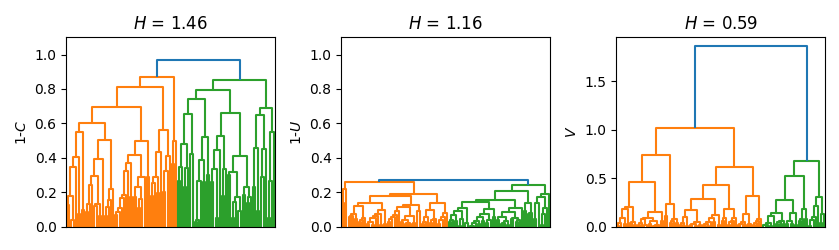}\\
     \hspace{.5cm} (a) \hspace{3.5cm} (b) \hspace{3.8cm} (c)\\ \vspace{.5cm}
     \includegraphics[width=1 \textwidth]{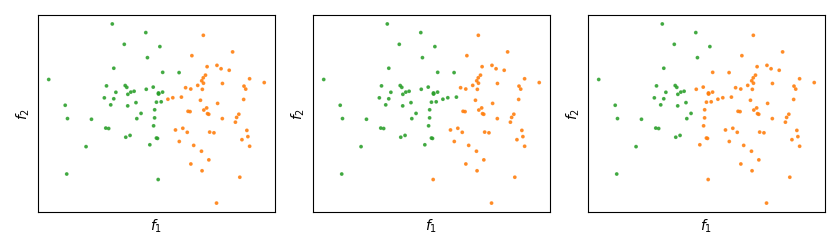}\\
     \hspace{.5cm} (d) \hspace{3.5cm} (e) \hspace{3.8cm} (f)
 \caption{Illustration of the agglomerative clustering by using the three considered approaches respectively to a specific dataset following the cluster structure shown in Fig.~\ref{fig:cluster}(a). The homogeneity index $H$ indicates that the proportional and uniform methods have the greatest overall homogeneity, while the Ward's approach is characterized by heterogeneity along all its hierarchical levels.}\label{fig:example_case1}
\end{figure}

\begin{figure}
  \centering
     \includegraphics[width=1 \textwidth]{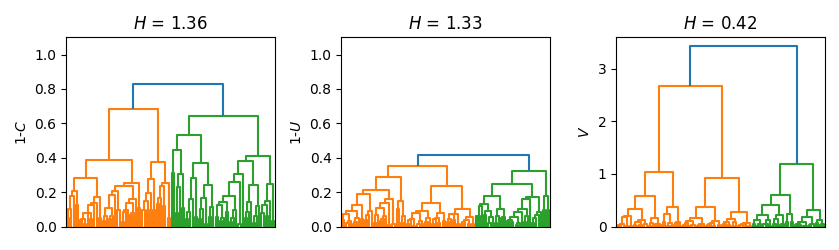}\\
     \hspace{.5cm} (a) \hspace{3.5cm} (b) \hspace{3.8cm} (c)\\ \vspace{.5cm}
     \includegraphics[width=1 \textwidth]{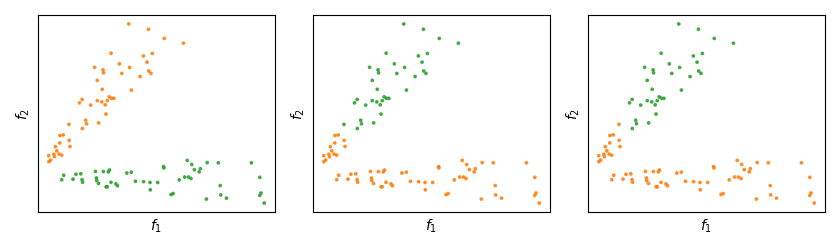}\\
     \hspace{.5cm} (d) \hspace{3.5cm} (e) \hspace{3.8cm} (f)
 \caption{Illustration of the agglomerative clustering by using the three considered approaches respectively to a specific dataset following the cluster structure shown in Fig.~\ref{fig:cluster}(b). Larger homogeneity  values have again been obtained in the case of the proportional and uniform clustering approaches.}\label{fig:example_case2}
\end{figure}

\begin{figure}
  \centering
     \includegraphics[width=1 \textwidth]{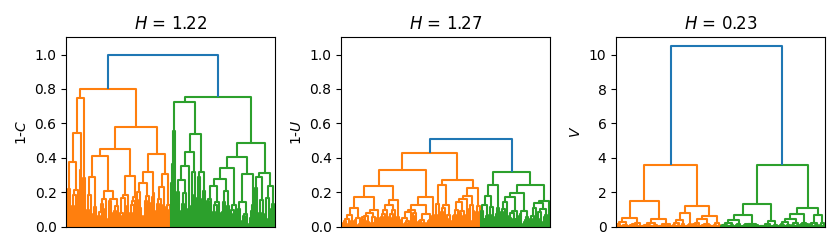}\\
     \hspace{.5cm} (a) \hspace{3.5cm} (b) \hspace{3.8cm} (c)\\ \vspace{.5cm}
     \includegraphics[width=1 \textwidth]{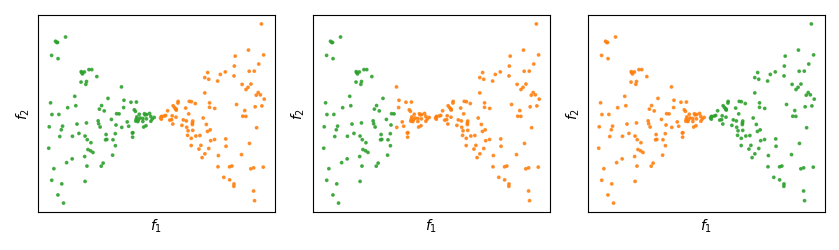}\\
     \hspace{.5cm} (d) \hspace{3.5cm} (e) \hspace{3.8cm} (f)
 \caption{Illustration of the agglomerative clustering by using the three considered approaches respectively to a specific dataset following the cluster structure shown in Fig.~\ref{fig:cluster}(c).  The largest homogeneity has again been obtained by using the proportional and uniform clustering methods.}\label{fig:example_case3}
\end{figure}

Figures~\ref{fig:example_case1},\ref{fig:example_case2}, and \ref{fig:example_case3} illustrate the results obtained for each of the three configurations described above for specific parameter configurations.

It should be kept in mind that the obtained results are specific not only to the adopted types of clustering structures (and respectively chosen parameters and number of points), but also to the parameters considered for each of the methods. Additional studies would be needed to be performed in the case of other types of clusters and configurations.

\begin{figure}
  \centering
     \includegraphics[width=.6 \textwidth]{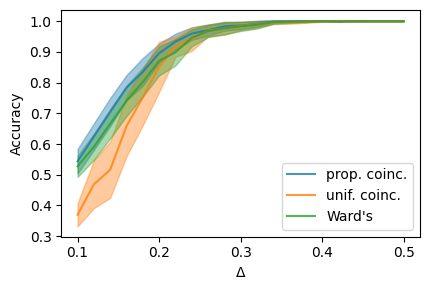}
 \caption{Average $\pm$ standard deviation of the accuracy index obtained for $T=100$ instances of the cluster configuration shown in Fig.~\ref{fig:cluster}(a). The proportional coincidence and Ward's methodologies yielded similar accuracies, followed by the uniform coincidence approach.}\label{fig:case1}
\end{figure}

Figure~\ref{fig:case1} depicts the average $\pm$ standard deviation of the performance index $\alpha$ obtained in the case of the clustering structure shown in Figure~\ref{fig:cluster}(a). The proportional and Ward's agglomerative approaches led to the best identification of the two original clusters.

\begin{figure}
  \centering
     \includegraphics[width=.6 \textwidth]{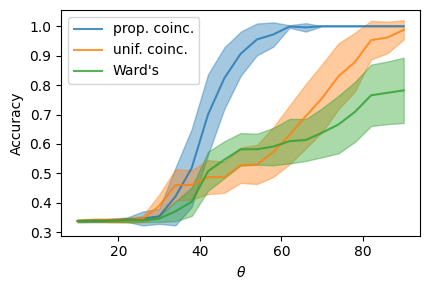}
 \caption{Average $\pm$ standard deviation of the accuracy index obtained for $T=100$ instances of the cluster configuration shown in Fig.~\ref{fig:cluster}(b). The proportional coincidence allowed the highest accuracy index values for the majority of the instances $\theta$. Ward's method yielded higher accuracy values than the uniform coincidence for $40 < \theta < 60$ degrees, while the uniform coincidence allowed higher accuracy for $\theta < 40$ and $\theta > 60$ degrees.}\label{fig:case2}
\end{figure}

The results obtained for the second considered clustering structure, shown in Figure~\ref{fig:cluster}(b), are presented in Figure~\ref{fig:case2}. The proportional agglomerative method led to the best results for most of the considered values of $\theta$.

\begin{figure}
  \centering
     \includegraphics[width=.6 \textwidth]{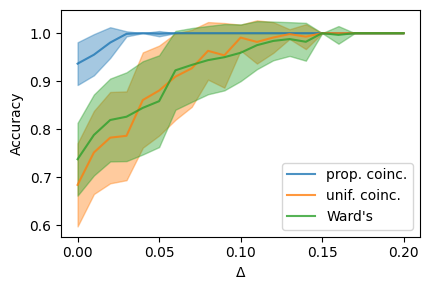}
 \caption{Average $\pm$ standard deviation of the accuracy index obtained for $T=100$ instances of the cluster configuration shown in Fig.~\ref{fig:cluster}(c). The proportional coincidence allowed the highest accuracy index values for the majority of the instances $\Delta$, with the two other approaches yielding similar accuracies.}\label{fig:case3}
\end{figure}

Figure~\ref{fig:case3} illustrates the performance results obtained in the case of the clustering structure shown in Figure~\ref{fig:cluster}(c). The best results were obtained by using the proportional agglomerative approach for all values of $\Delta$.

\begin{figure}
  \centering
     \includegraphics[width=.6 \textwidth]{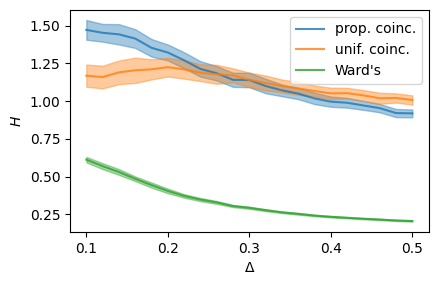}
 \caption{The values of the homogeneity index $H$ in terms of the parameter $\Delta$ obtained for the dataset involving two adjacent normal distributions.}\label{fig:case1_H}
\end{figure}

\begin{figure}
  \centering
     \includegraphics[width=.6 \textwidth]{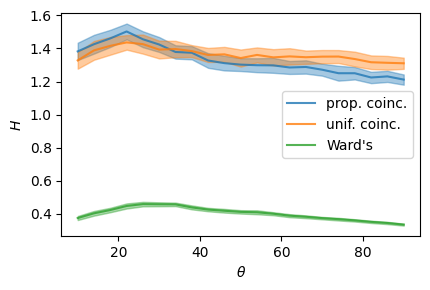}
 \caption{The values of the homogeneity index $H$ in terms of the parameter $\theta$ obtained for the dataset involving two clusters with points uniformly distributed within sections of a disk, separated by angle $\theta$.}\label{fig:case2_H}
\end{figure}

\begin{figure}
  \centering
     \includegraphics[width=.6 \textwidth]{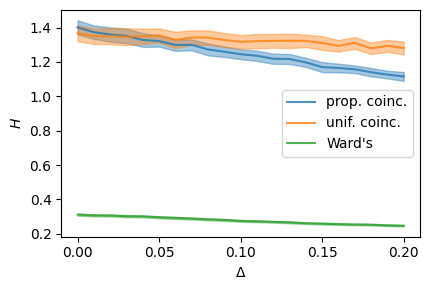}
 \caption{The values of the homogeneity index $H$ in terms of the parameter $\Delta$ obtained for the dataset involving two clusters with points uniformly distributed within sections of a disk separated by a distance $\Delta$.}\label{fig:case3_H}
\end{figure}

The values of the homogeneity index $H$ obtained in terms of respective parameter variations in the three above considered experiments are presented in Figures~\ref{fig:case1_H},~\ref{fig:case2_H}, and ~\ref{fig:case3_H}. In all cases, the Ward's methodology was characterized by substantially smaller overall homogeneity, despite the fact that the distribution of points within the two main clusters was originally uniform.

\section{Concluding Remarks}

The subject of data clustering has constituted an endeavor of great relevance, as it provides the basis not only for data analysis, but also to artificial intelligence.  The importance of this area has been substantiated by the impressive number of related publications and applications to an ever increasing number of problems and tasks.
Among the several approaches that have been considered for data clustering, the family of methods known as hierarchical agglomerative clustering presents special relevance because it does not require pre-specification of the number of clusters and also allows an effective identification of the relationships among the involved subclusters, as revealed by the respectively obtained dendrograms.

Among the several existing agglomerative hierarchical clustering approaches, the Ward's method has received special attention as a consequence of its ability to identify clusters.  Founded on variance minimization, this approach is intrinsically related to clusters following the normal distribution, which include circularly symmetric and elongated clusters, which also exhibit symmetry.  At the same time, the Ward's approach has been found to present a tendency to identify clustering and subclustering structures even when there are none, as is the case in uniform noise.

While traditional agglomerative hierarchical clustering methods have been based on the concept of distance or data variance, the possibility to use similarity has been little investigated.  The present work was aimed at proposing two agglomerative hierarchical clustering approaches based on multiset concepts --- including multiset cardinality, union and intersection. The two described methods are aimed at dealing with clustering data in uniform and proportional spaces, with the latter approach corresponding to the recently presented concept of coincidence similarity.  In addition to describing these two methods, the present work also presented a comparison of relative performance of them and the traditional Ward's approach respectively to three types of clustering structures.

In order to provide context to the described methods and experiments, this work started by presenting and discussed the important and not often addressed topics of uniform and proportional data spaces, followed by a mostly self-contained description of multiset concepts and respectively derived similarity indices, including the Jaccard, interiority and coincidence indices.  It was then shown that these indices intrinsically perform proportional comparison between data elements.  Of particular relevance, special attention has been given to approaching cluster detection in uniform and proportional spaces based on respective cluster models, which were then considered in the reported experiments.  The measured employed to quantify the clustering performance, which is itself based on the coincidence similarity, was also defined and illustrated. Two indices have also been described and employed in order to: (i) provide indication about the overall homogeneity of a given dendrogram; and (ii) to quantify the relevance of each obtained dendrogram branch (group) as a possible cluster in the original data set.  The respectively obtained values contributed substantially for estimating and comparing the properties of the adopted methods respectively to the considered data sets.

Several interesting results have been described and discussed, with substantial potential for impacting research in agglomerative hierarchical clustering.  First, it was preliminary verified that the uniform and proportional clustering methods are intrinsically suitable for dealing with data underlain by uniform and proportional clusters, with the Ward's methodology not only being intrinsically unable to cope with clusters in proportional spaces, but also presenting a strong tendency to identifying non-existent subclustering structures. Both the uniform and proportional clustering methods were found, at least for the considered type of data, to be particularly robust concerning the identification of non-existent subclustering structures.

A subsequent systematic analysis was then reported which involved three distinct types of clusters, namely with points distributed according to circularly symmetric normal distribution, as well as two situations involving points uniformly distributed within a section of a disk.  All these situations involved uniform feature spaces.  The obtained results suggest that, at least for this type of data and parameter configurations, the proportional agglomerative clustering approach based on the coincidence similarity index turned out to be capable of outperforming the two other hierarchical methods (namely uniform and Ward's) in most situations, allowing substantially better performance in the two cases involving clusters not following the normal distribution.  

The obtained results substantiate the great potential of hierarchical clustering methods based on multiset similarity indices, especially the coincidence similarity index.  In particular, this approach was found to provide effective performance not only for clusters in proportional feature spaces, to which it is especially suited, but also to several types of clusters in uniform feature spaces.  These findings have substantial implications not only from the theoretical point of view, with possible applications to neuronal networks and deep learning, but also constituting an interesting alternative to be considered for a variety of applications in pattern recognition and artificial intelligence. Another particularly relevant contribution of the present work has been the description and characterization of uniform and proportional spaces, as well as of respective comparison indices.

Several further developments are motivated by the presented concepts, methods, and results. These include the consideration of more than two clusters, other types of clusters, data dimensions larger than two, as well as the presence of noise, outliers, and interference in the data set.  Other interesting prospects include the application of the described concepts and methods to other problems in pattern recognition and artificial intelligence, including similarity based approaches to hierarchical deep learning.

\section*{Acknowledgments}
Alexandre Benatti thanks MCTI PPI-SOFTEX (TIC 13 DOU 01245.010\\222/2022-44). Luciano da F. Costa thanks CNPq (grants no.~307085/2018-0 and 313505/2023-3) and FAPESP (grant 2022/15304-4).

\bibliography{ref}

\begin{thebibliography}{10}

\bibitem{da2019modeling}
L.~da~F.~Costa.
\newblock Modeling: The human approach to science ({CDT}-8).
\newblock \url{https://www.researchgate.net/publication/333389500_Modeling_The_Human_Approach_to_Science_CDT-8}, 2019.

\bibitem{fukunaga2013}
K.~Fukunaga.
\newblock {\em Introduction to statistical pattern recognition}.
\newblock Elsevier, 2013.

\bibitem{bishop2006}
C.~M. Bishop and N.~M. Nasrabadi.
\newblock {\em Pattern recognition and machine learning}, volume~4.
\newblock Springer, 2006.

\bibitem{duda2000pattern}
R.~O. Duda, P.~E. Hart, and D.~G. Stork.
\newblock {\em Pattern classification}.
\newblock Wiley-Interscience New York, 2nd edition, 2000.

\bibitem{da2018shape}
L.~da~F.~Costa and R.~M. Cesar~Jr.
\newblock {\em Shape classification and analysis: theory and practice}.
\newblock Crc Press, 2018.

\bibitem{amancio2014systematic}
D.~R. Amancio, C.~H. Comin, D.~Casanova, G.~Travieso, O.~M. Bruno, F.~A. Rodrigues, and L.~da~F.~Costa.
\newblock A systematic comparison of supervised classifiers.
\newblock {\em PloS One}, 9(4):e94137, 2014.

\bibitem{rodriguez2019clustering}
M.~Z. Rodriguez, C.~H. Comin, D.~Casanova, O.~M. Bruno, D.~R. Amancio, L.~da~F. Costa, and F.~A. Rodrigues.
\newblock Clustering algorithms: A comparative approach.
\newblock {\em PloS One}, 14(1):e0210236, 2019.

\bibitem{ward1963hierarchical}
J.~H. Ward~Jr.
\newblock Hierarchical grouping to optimize an objective function.
\newblock {\em Journal of the American statistical association}, 58(301):236--244, 1963.

\bibitem{murtagh2012}
F.~Murtagh and P.~Contreras.
\newblock Algorithms for hierarchical clustering: an overview.
\newblock {\em Wiley Interdisciplinary Reviews: Data Mining and Knowledge Discovery}, 2(1):86--97, 2012.

\bibitem{da2021further}
L.~da~F.~Costa.
\newblock Further generalizations of the {J}accard index.
\newblock \url{https://www.researchgate.net/publication/355381945_Further_Generalizations_of_the_Jaccard_Index}, 2021.

\bibitem{costa2022simil}
L.~da~F.~Costa.
\newblock On similarity.
\newblock {\em Physica A: Statistical Mechanics and its Applications}, 599:127456, 2022.

\bibitem{da2022coincidence}
L.~da~F.~Costa.
\newblock Coincidence complex networks.
\newblock {\em Journal of Physics: complexity}, 3(1):015012, 2022.

\bibitem{costa2023mneurons}
L.~da~F.~Costa.
\newblock Multiset neurons.
\newblock {\em Physica A: Statistical Mechanics and its Applications}, 609:128318, 2023.

\bibitem{da2021real}
L.~da~F.~Costa.
\newblock Real-valued {J}accard and coincidence based hierarchical clustering.
\newblock {\em ResearchGate}, 2021.

\bibitem{JohnsonWichern}
R.~A. Johnson and D.~W. Wichern.
\newblock {\em Applied Multivariate Statistical Analysis}.
\newblock Pearson, 2015.

\bibitem{Anderson}
T.~W. Anderson.
\newblock {\em An Introduction to Multivariate Statistics Analysis}.
\newblock Wiley-Interscience, 3rd edition, 2000.

\bibitem{mcculloch}
W.~S. McCulloch and W.~Pitts.
\newblock A logical calculus of the ideas immanent in nervous activity.
\newblock {\em The bulletin of mathematical biophysics}, 5:115--133, 1943.

\bibitem{lecun2015deep}
Y.~LeCun, Y.~Bengio, and G.~Hinton.
\newblock Deep learning.
\newblock {\em Nature}, 521(7553):436--444, 2015.

\bibitem{pouyanfar2018survey}
S.~Pouyanfar, S.~Sadiq, Y.~Yan, H.~Tian, Y.~Tao, M.~P. Reyes, M.-L. Shyu, S.-C. Chen, and S.~S. Iyengar.
\newblock A survey on deep learning: Algorithms, techniques, and applications.
\newblock {\em ACM Computing Surveys (CSUR)}, 51(5):1--36, 2018.

\bibitem{Jac:wiki}
Wikipedia.
\newblock Jaccard index, 2021.
\newblock \url{https://en.wikipedia.org/wiki/Jaccard_index}. [Online; accessed 10-Oct-2021].

\bibitem{fisher1970statistical}
R.~A. Fisher.
\newblock Statistical methods for research workers.
\newblock In {\em Breakthroughs in statistics: Methodology and distribution}, pages 66--70. Springer, 1970.

\bibitem{bonett2000sample}
D.~G. Bonett and T.~A. Wright.
\newblock Sample size requirements for estimating pearson, kendall and spearman correlations.
\newblock {\em Psychometrika}, 65:23--28, 2000.

\bibitem{kim2015instability}
Y.~Kim, T.-H. Kim, and T.~Erg{\"u}n.
\newblock The instability of the pearson correlation coefficient in the presence of coincidental outliers.
\newblock {\em Finance Research Letters}, 13:243--257, 2015.

\bibitem{da2022brief}
L.~da~F.~Costa.
\newblock A brief guide to the coincidence similarity and its applications.
\newblock \url{https://www.researchgate.net/publication/362713778_A_Brief_Guide_to_the_Coincidence_Similarity_and_its_Applications}, 2022.

\bibitem{costa2021comparing}
L.~da~F.~Costa.
\newblock Comparing cross correlation-based similarities.
\newblock {\em arXiv preprint arXiv:2111.08513}, 2021.

\bibitem{vijaymeena2016a}
M.~K. Vijaymeena and K.~Kavitha.
\newblock A survey on similarity measures in text mining.
\newblock {\em Machine Learning and Applications: An International Journal}, 3(2):19--28, 2016.

\bibitem{da2021multisets}
L.~da~F.~Costa.
\newblock Multisets.
\newblock \url{https://www.researchgate.net/publication/355437006_Multisets}, 2021.

\bibitem{benatti2022neuromorphic}
A.~Benatti, H.~F. de~Arruda, and L.~da~F. Costa.
\newblock Neuromorphic networks as revealed by features similarity.
\newblock {\em arXiv preprint arXiv:2207.10571}, 2022.

\bibitem{domingues2022identification}
G.~S. Domingues, E.~K. Tokuda, and L.~da~F.~Costa.
\newblock Identification of city motifs: a method based on modularity and similarity between hierarchical features of urban networks.
\newblock {\em Journal of Physics: Complexity}, 3(4):045003, 2022.

\bibitem{tokuda2023cross}
E.~K. Tokuda, R.~Lambiotte, and .~da~F.~Costa.
\newblock Cross-relation characterization of knowledge networks.
\newblock {\em The European Physical Journal B}, 96(11):144, 2023.

\bibitem{benatti2023quantifying}
A.~Benatti, A.~C.~M. Brito, D.~R. Amancio, and L.~da~F.~Costa.
\newblock Quantifying the hierarchical adherence of modular documents.
\newblock {\em Journal of Physics: Complexity}, 4(4):045008, 2023.

\bibitem{benatti2024simple}
A.~Benatti and L.~da~F.~Costa.
\newblock Simple games on complex networks.
\newblock \url{https://www.researchgate.net/publication/380457853_Simple_Games_on_Complex_Networks}, 2024.

\bibitem{benatti2023multilayer}
A.~Benatti and L.~da~F.~Costa.
\newblock Multilayer multiset neuronal networks--mmnns.
\newblock {\em arXiv preprint arXiv:2308.14541}, 2023.

\bibitem{benatti2023two}
A.~Benatti and L.~da~F.~Costa.
\newblock Two approaches to supervised image segmentation.
\newblock {\em arXiv preprint arXiv:2307.10123}, 2023.

\bibitem{costa2023mulsetions}
L.~da~F.~Costa.
\newblock Mulsetions and intervalions: Multiset generalizations of functions.
\newblock \url{https://hal.science/hal-04201139v2/file/intervalions.pdf}, 2023.

\bibitem{tokuda2022}
E.~K Tokuda, C.~H. Comin, and L.~da~F. Costa.
\newblock Revisiting agglomerative clustering.
\newblock {\em Physica A: Statistical mechanics and its applications}, 585:126433, 2022.

\end{thebibliography}
\bibliographystyle{unsrt}

\end{document}